Short Paper

# Web-based Management Information System of Cases Filed with the National Labor Relations Commission

Aaron Paul M. Dela Rosa
College of Information and Communications Technology, Bulacan State University, Philippines
aaronpaul.delarosa@bulsu.edu.ph

**Abstract**

*Purpose* – This study was developed to describe the daily operations and encountered problems of the National Labor Relations Commission Regional Arbitration Branch No. IV (NLRC RAB IV) through conducted observations and interviews. These problems were addressed and analyzed to be the features of the developed web-based management information system (MIS) for cases.

*Method* – The research methodology utilized in this project was the descriptive developmental approach. The Agile Software Development methodology was followed to develop the system. It was used to quickly produce the desired output while allowing the user to go back through phases without finishing the whole cycle.

*Results* – The system covered managing filed complaints, Single-Entry Approach (SEnA), labor cases, and report generation. The findings, through the interview, of handling records were inconsistent and inaccurate. This study also focused on ensuring the Data Privacy Act of 2012, protecting the database's information using the XOR Cipher Algorithm. This study was evaluated using standard web evaluation criteria. Using the criteria, the study's overall mean was 4.27 and 4.43, with the descriptive meaning of *Very Good*, which showed that the system was accepted as perceived by experts and end-users, respectively.

*Conclusion* – Management of filed cases is a vital process for the Commission. With that said, developing the web-based management information system could ease the internal operations of handling and managing filed labor cases. Moreover, respondents and complainants can easily determine their filed cases' status using the case status tracking system.

*Recommendations* – For further improvements to the system, additional printable documents may be added that could be found needed by the Commission. Moreover, completion of the software development methodology may be done to complete the cycle. Lastly, further research about the effectiveness of the web-based system may be conducted for further enhancements of the system.

*Research Implications* – Services of the Commission may be enhanced upon the implementation of the developed web-based management information system of filed cases. Additionally, offices will have ease of access with the provision of different user levels of access.

*Keywords* – management information system, information technology, information systems, agile software development methodology, labor relations

## INTRODUCTION

The utilization of web technologies in today's era gives more ways of sharing information through the internet and local connections. The development of web information systems (WIS) is rising globally, making businesses and companies more competitive (Hanks, 2018; Schewe & Thalheim, 2019). Web technologies offer a wide range of information-sharing, whether online or offline. Offline sharing of information through local area networks can be maximized by WIS in sharing locally connected devices. Online services from WIS allow users from different locations to access or manipulate information.

Usage of WIS has its advantages (Hanks, 2018), which is why even government agencies are investing in such information systems. WIS allows clients far from the organization or company to connect, allowing them to communicate their needs and purpose. WIS also provides the advantages of transacting online, allowing clients to register information through web portals of government agencies. Several government agencies are currently enjoying the services offered by WIS. The Department of Foreign Affairs (DFA) is using WIS to allow clients to register for a passport application and renewal, providing the clients only visit their office for the purpose of taking pictures for the passport (Department of Foreign Affairs, n.d.). This lessens the time consumed by the people queuing at their offices. This also helps the clients to look for available schedules online for passport renewals without the effort of visiting the office and knowing that there are no available schedules anymore. Another government agency using WIS is the Department of Tourism (DoT) (Department of Tourism, 2021). DoT shares, through WIS, the beauty of the Philippine islands. This helps promote the local economy and provides opportunities for foreigners to visit and see the beauty of the country. This allows the Philippines to be recognized and honored by foreign countries. Recently, DoT has been recognized and cited by Forbes.com as one of the hottest spots in Asia.

The National Labor Relations Commission (NLRC), a government agency under the Department of Labor and Employment (DOLE), is a quasi-judicial court that handles labor cases filed by complainants against its respondents with a specific type of jurisdiction of labor arbiters. In the Philippines, according to the official website of NLRC (National Labor Relations Commission, n.d.), 31,785.2 cases were received, and 31,885.4 cases were disposed of by the regional arbitration branches (RABs) of NLRC from the years 2001 to 2010. These cases received and disposed of by the NLRC fall under the jurisdiction of the labor arbiters. These jurisdictions of labor arbiters are unfair labor practice cases, termination disputes, cases involving wages, rates of pay, hours of work, money claims, and other cases as may be provided by law. These jurisdictions are mentioned in Article 217 of Presidential Decree No. 442, otherwise known as the Labor Code of the Philippines.

NLRC has many branches in the whole country. One of the biggest RABs is Region 4, covering both Region 4A, which includes Cavite, Laguna, Batangas, Rizal, and Quezon (CALABARZON), and Region 4B, which includes Mindoro, Marinduque, Romblon, and Palawan (MIMAROPA). NLRC RAB IV is located in Calamba, Laguna, a few meters away from the city's municipal hall. It has eight offices to cater to the labor arbiters (LA), including the office of the executive labor arbiter (ELA). All offices have labor arbitration associates (LAA). These are the people allowed to do the tasks of a labor arbiter if the labor arbiter is not present during the finalization of a filed case. Each labor arbiter handles several cases each day, around 2,000 cases a year, including Palawan cases, making their offices filled with filed complaints and crowding their offices with spaces occupied by the papers. With this, the branch even had a room just for the documents, and each office of the labor arbiters is still filled with such documents. A need for a better record management system arises from the number of cases received and handled by the RAB.

Some employees of the agency visit Palawan, including the ELA, one of the farthest places covered by Region 4, to accommodate filed cases at the said location, doing this every other month. People complained that Palawan only allowed them to ask about the progress of their filed complaints every other month, making them wait for a month or two to know what had happened in their case.

The development of a management information system for cases for the NLRC RAB IV helped reduce paper work (Hanks, 2018) and store information in a centralized database (Yang, Xiong, & Ren, 2020). Each labor arbitrator handles different sets of cases, making the centralization of the database per office of the labor arbitrator unnecessary. Also, with this, only the employees of each office have the right to log in to the system to fully use the management information system. Moreover, the system has a labor case status tracking system that allows complainants or respondents to view the status of their labor cases online. This helped the complainant or respondent to lessen the effort and time of visiting the office for a case status inquiry. The system also has a report generation needed by each labor arbitrator's office to be submitted to the main office of the NLRC. This reduced the use of spreadsheets, which sometimes cause data inconsistency.

## LITERATURE REVIEW

Lee-Geiller and Lee (2019) studied the importance of government websites as a medium of communication for the public sector. The authors highlighted the use of technology and its involvement in the advancement of services and policies in the government sector. They also pointed out how such systems can influence the public to reach out and communicate with the government.

The provision of online access to information in the web-based management information system provided ease of access to any or both parties involved to search and know the progress or the status of the filed labor case. This does provide quick access to information that would also lessen the burden of any or both parties by visiting the agency to ask and know the progress of their labor case. This aids the public in reaching out to the agency when an inquiry is needed.

According to a local study conducted by Bautista, Calimpusan, David, and Delos Reyes (2012), "Management Information Systems are not only computer systems—these systems encompass three primary components: technology; people (individuals, groups, or organizations); and data/information for decision making." Bautista et al. (2012) aimed to provide better file management of police suspect profiles and records related to such.

As the developed system aimed to manage SEnA and filed labor cases, proper data management on the developed system was provided for the users to maximize their time in searching for the records of the filed labor cases. Modifications to the system shall be done with the approval of the labor arbiter in the designated office. This assures that the data will keep its consistency from the printed documents to the data stored in the web-based management information system.

The use of web technologies in developing the web-based management information system was maximized to allow the employees of NLRC RAB IV to provide quality services by providing the complainants and the respondents with the necessary data they need as quickly as possible. The developed system allowed the user to view the status of the filed labor case, which shortened the time of searching for the necessary document in the pile of filed labor cases. The developed system also allowed quality management of data, from storage, retrieval, and modification. With this quality of data management, the end-users of the web-based system will have the assurance that the data presented to them will be consistent. With data being consistent throughout the processing of a labor case, the reports to be submitted to the main office are valid and acceptable.

Handling labor cases has multiple things that need to be considered. Xie (2015) once said that handling or managing labor inspections should be done to correct the violations committed by the employers to their employees, following the procedure of the government to allow proper decisions based on the jurisdiction of the lawyer administering the case.

Administering arbitration from other countries in comparison with the Philippines is not that different. The NLRC administers filed arbitrations in their office in accordance with the provided jurisdiction of labor arbiters by the government. With such a provision, labor arbiters will know their position in deciding an arbitration to amicably settle the violation committed by the respondent to the complainant.

Tarasewicz and Borofsky (2013) concluded that "determining how employment issues are resolved is an important matter for each country." This study is meant to present the European perspective of legislation and arbitration of labor law in European countries, specifically focusing more on French labor and arbitration law. Employment issues, or labor issues, are concerned with different factors of a person as an employee of an organization. Employees of an organization or an agency are treated as subordinate to their employers. Employee and employer relationships are treated differently in dependence on the country's governing rules of arbitration and legislation.

Katz, Kochan, and Colvin (2015) identified the processes by which labor, management, and the government interact with a labor law arbitration. As mentioned in the study by Katz et al. (2015), labor pertains to the employees of the company, management pertains to the employer, and the government pertains to national, regional, or local agencies that deal with labor relations. Also, not just identifying the processes, Katz et al. (2015) also considered the historical events of a country's affect the government policy on how to deal with unfair labor practices committed by private sector employees.

## METHODOLOGY

A descriptive method was applied to this study as it described the daily operations of NLRC RAB IV, where the work of managing filed cases was digitized using the developed system. Information gathered in analyzing human behavior in accepting and adapting to change with the descriptive method was the best approach. It aims to be more familiar with a specified topic and focuses on an area of interest that naturally occurs in the environment of the intended subject, or in this study, the end-users. Observations were conducted on the daily operations of NLRC RAB IV that determined the current state of the system and technology used by the agency. An interview was conducted with the executive labor arbiter, from which additional information was gathered and assurance of the observed data was validated. With this collected data, the researcher also identified the problems encountered in the daily operations of NLRC RAB IV (Mortensen, 2019). For each of the stated problems, suggestions were provided on how to improve the operations of the Commission to be part of the features of the developed system.

Developmental research was applied to the consistency, accuracy, and effectiveness of the developed system. In this type of research, the processes were studied systematically, developed, and evaluated to meet the set criteria for users' acceptance of the developed system. After gathering data and additional information through observations and interviews, after going through the descriptive method process, then, developmental research was adopted. The gathered data on daily operations and problems encountered by the client was analyzed, which led to the development of the web-based management information system for labor cases, which improved the daily operations of handling filed cases.

## *Project Development*

The system was developed by following and applying the processes and phases of the System Development Life Cycle (SDLC). The developed system focused on using Agile software development methodology as its primary SDLC method (Leau, Loo, Tham, & Tan, 2012). This enabled developing the web-based system by going back and forth on different phases, revisiting a phase if there are changes in the latter phases, and modifying contents from past phases if there are inconsistencies or anomalies that happened during the development of the system. Agile software development is suitable for developing fast-paced systems as it allows quick iteration of its phases and it also allows revisiting past phases if there are modifications to be made.

*Requirements* The data gathering using interviews and survey questionnaires was done in this phase. The purpose of the interview and survey questionnaires was to identify the daily operations and problems encountered by NLRC RAB IV. With this gathered data, planning of features and proposal of a web-based management information system for labor cases to the Commission was able to be done.

*Plan.* In this phase, the possible features and functionalities included in the web-based system were identified. With the help of the gathered data, the proposal of the possible features and functionalities of the web-based systems was made to the NLRC.

*Design.* A database was designed using an entity relationship diagram that serves as the storage of complaints, SEnA, and filed labor cases. In this phase, a user interface was also designed that fits the requirements. A data flow diagram was used to determine how each data flowed and which functionalities each user type should have access to.

*Develop.* After the finalization of the required elements in the design phase, the web-based management information system, using and maximizing web technologies, was developed. In this phase, all the features of the developed system were delivered. As part of the development of the system, debugging of the system was also done. Debugging assures that the developed system operates properly and correctly, as debugging allows the correction of all errors encountered during the development of the system. This phase will also include user acceptance testing. All user requirements were verified and validated, and modifications to be made were identified before the implementation of the developed system. Different development tools were used to achieve the development of the web-based management information system. Used to develop both the front-end and back-end of the web-based system were Hypertext Markup Language (HTML), Cascading Style Sheet (CSS), JavaScript (JS), and Hypertext Preprocessor (PHP). A framework, such as jQuery, was also used as a development tool. My Structured Query Language (MySQL) was used as the database of the system where all data is and will be stored to be used by the web-based system.

*Released, tracked, and monitored.* In the release phase, the system is deployed and will be operational for the agency and used by its intended users. After the deployment, the tracking and monitoring phase will be completed for the maintenance of the system. The study no longer covers these phases as the study only covers up to the testing of the developed web-based management information system.

## *System Evaluation: Standard Web Evaluation Criteria*

The respondents to the developed system were evaluated for acceptability. This helped to identify the needs and satisfaction of the Commission and its employees in the developed web-based management information system for labor cases. According to the case flow at the regional arbitration branch level, SEnA and complaint officers are the first to process and gather the information of the complainants and the respondents before proceeding to an actual case handled by the labor arbiters.

Expert or judgment sampling (Glen, 2015) was used on 30 IT experts and professionals that assessed the acceptability of the developed system. Experts' responses were included as part of the responses that provided evaluations of the developed system's technicalities. On the other hand, convenience sampling (Maheshwari, 2017) was used on the 30 NLRC RAB IV employees. These employees were the end-users and are categorized as labor arbiters, including labor arbitration associates and the executive labor arbiter, SEnA officers, and complaint officers.

Surveys were conducted with the employees of NLRC RAB IV, including labor arbiters, labor arbitration associates, SEnA officers, and complaint officers, to assess the level of acceptability of the developed web-based management information system for labor cases. To assess the acceptability of the developed system, standard web-evaluation criteria were used. They were computed using the five-point Likert-type Scale as shown in Table 1.

Table 1. Five-point Likert Scale

| Scale | Range | Descriptive Rating |
|---|---|---|
| 5 | 4.50 – 5.00 | Excellent |
| 4 | 3.50 – 4.49 | Very Good |
| 3 | 2.50 – 3.49 | Good |
| 2 | 1.50 – 2.49 | Fair |
| 1 | 1.00 – 1.49 | Poor |

The standard web evaluation criteria were based on the models that were analyzed by Valaviius and Vipartien (2013), which formed three models that were used to evaluate a website, namely: (1) the quality model, (2) usability model, and (3) the satisfaction model. These models were compiled from other existing models that were compiled by Valaviius and Vipartien (2013) to develop a better website evaluation model. The quality model identifies the quality of the data presented in the system. This model was derived from the International Organization of Standardization (ISO) quality models (ISO 25000, ISO 2022). The usability model determines how the user will use and interact with the system. Appelman and Sundar (2019), Gómez, Caballero, and Sevillano (2014), and Sauro (2015) have incorporated the usability model into their research. The satisfaction model identifies the user's perception in terms of the presentation of system design and how well each component is connected to one another. The studies of Titiyal, Bhattacharya, and Thakkar (2019), Willis and Jozkowski (2019), and Yardley, Morrison, Bradbury, and Muller (2015) are a few of those that used the satisfaction model.

## RESULTS AND DISCUSSION

***The Common Problems being Encountered by the National Labor Relations Commission Regional Arbitration Branch No. IV in their Daily Operations***

The Executive Labor Arbiter (ELA) of the NLRC, Atty. Generoso V. Santos, described and discussed their daily operations during the interview. During the interview, when asked about the problems they encounter when sharing information on the case transition from SEnA to a proper labor case, ELA mentioned that sometimes the data is incomplete and inconsistent, specifically the information of the complainants and the respondent. The attention of the SEnA officer shall be called to clarify the details and information of both parties, including the nature of the complaint filed by the complainant.

Raffling from one labor arbiter to another is one of their significant problems. The information from the other office sometimes changes if re-shuffled to another labor arbiter. With that, the new administering labor arbiter is having difficulty assessing the case and needs the office's attention where the case is shuffled initially. Another problem encountered in sharing information is when both parties need to identify the status of their filed case. Any or both parties shall visit the Commission for a simple inquiry of their labor case's status. It burdens both parties to spend money, time, and effort just to learn such a small detail. These problems are the common problems they encounter, mostly in their daily operations.

*Development of the Functional Requirements of the Web-based Management Information System*

The requirements for the web-based management information system were drawn from the encountered problems in the Commission's daily operations that were gathered through the interview. Software Development Tools (SDTs) were utilized and maximized to develop the system. The developed system aimed to improve the daily operations of the Commission, specifically in managing records properly.

The system provided complainant and complaint management for complaint officers; SEnA management for SEnA officers; report generation for labor arbitration associates; and case management for labor arbiters. Generated reports from the system were provided in portable document format (PDF). This generation was done to view, print, or save the report for future use. Figure 1 shows the complaints folder page.

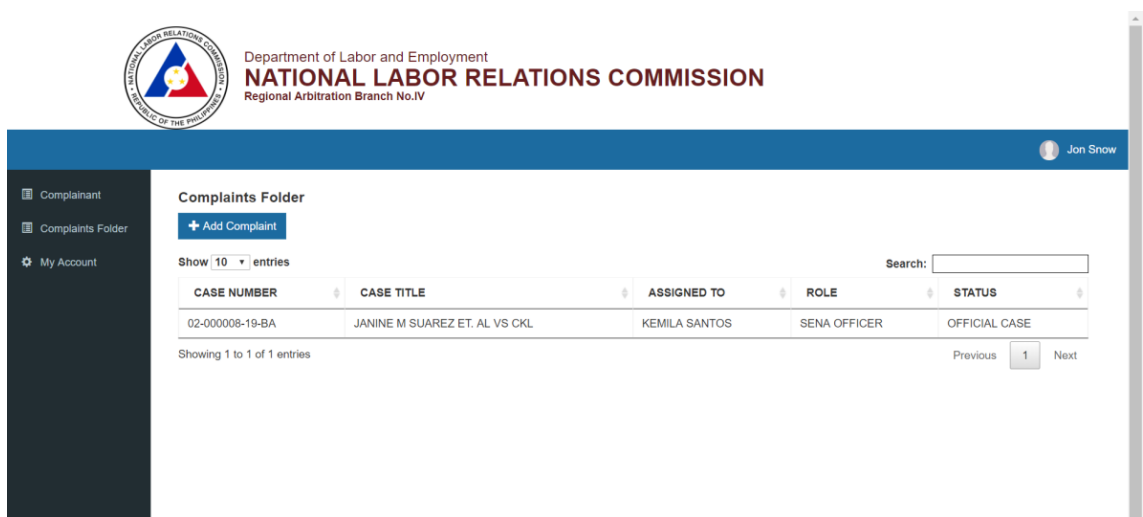

*Figure 1.* Complaints Folder page

The Complaints Folder provides the list of all filed complaints registered and stored by a complaint officer of the Commission. The Complaints Folder page displays all the complaints and to whom each complaint was assigned. The assignment of a complaint is the role of the complaint officer. The complaints folder page addressed the situation of gathering the complainant's information and the type of complaint they are filing against the respondent. This feature also addressed a better way of distributing a complaint to a SEnA officer by selecting an officer who will administer the SEnA and providing all gathered information upon transferring. Figure 2 shows the SEnA page.

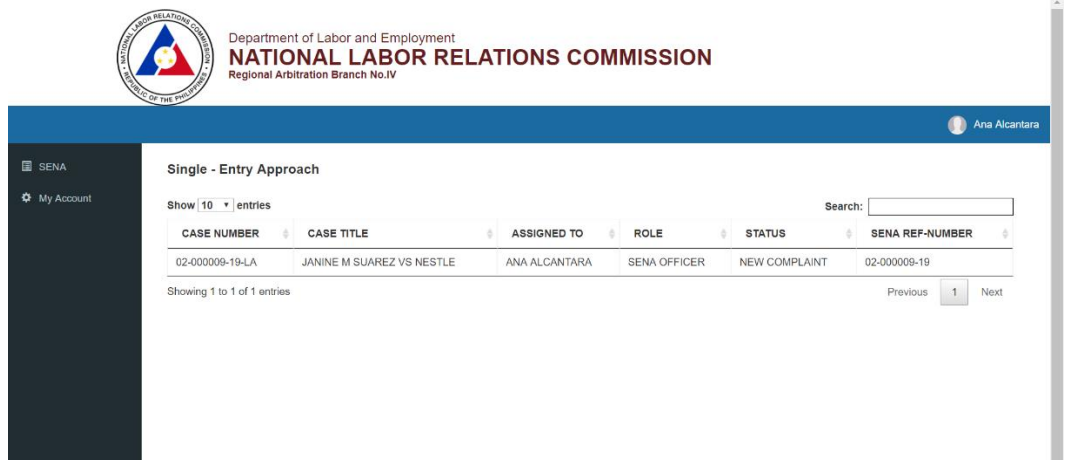

*Figure 2.* SEnA page

The SEnA page provides the list of all SEnA handled by the administering SEnA officer. The assignment of SEnA is the role of the complaint officer. This feature addressed the management of complaints that were transferred to a SEnA officer. It allowed the SEnA officer to manage all SEnA assigned by the complaint officer. Figure 3 shows the Case Maintenance page.

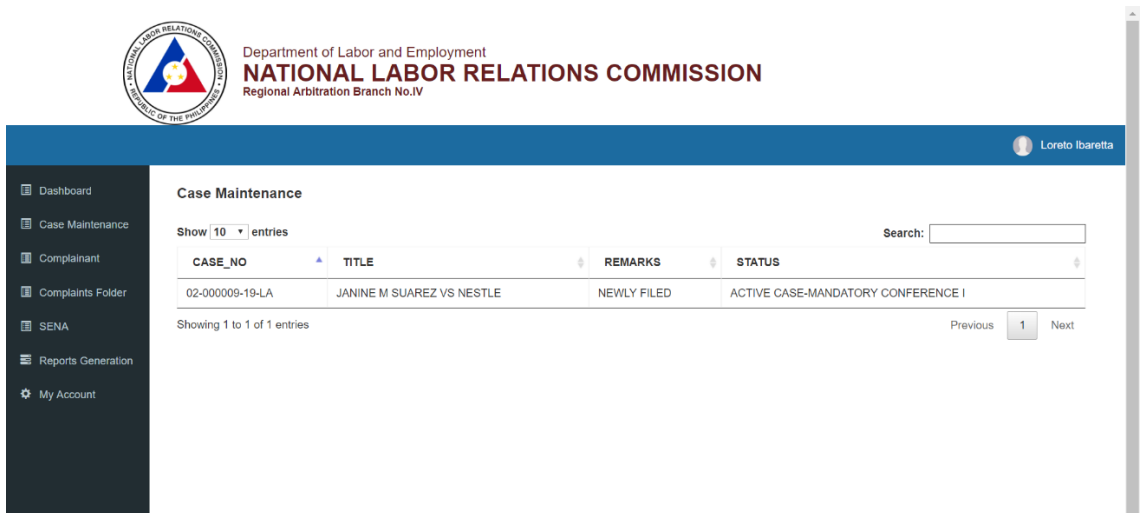

*Figure 3.* Case maintenance page

The Case Maintenance page provides the list of all cases and the logged user's actions on each of the filed case records. Cases are viewed and filtered by the office. That means cases that fall under the logged labor arbiter's office, or labor arbitration associate's office, are the displayed cases on the case maintenance page. The Case Maintenance page allows the labor arbiters to easily view all the details related to the filed labor cases they are handling. Case status could be changed from time to time by the labor arbitrator by providing details from the minutes of the past hearing meeting. Figure 4 shows a sample of a generated report in a PDF file.

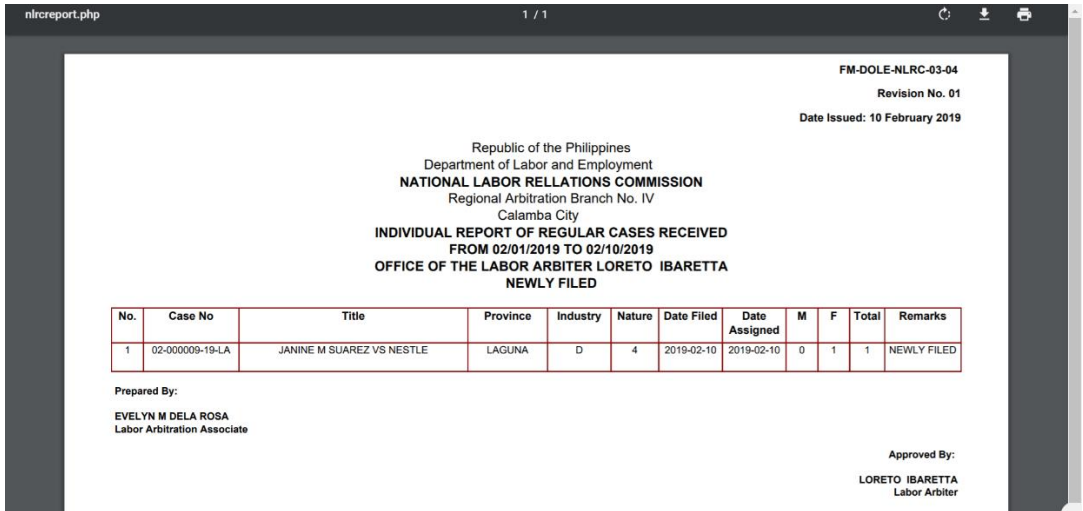

*Figure 4.* A generated report in PDF file format

The Reports Generation page provides access to generate necessary reports needed by the Commission in PDF format. Only the ELA, labor arbiters, and labor arbitration associates can generate reports. The logged user could choose a report type of regular or Overseas Filipino Workers (OFW) cases and choose a remark needed. As per the Commission's report on the main office, each report shall be by remark or status of the case. Figure 5 shows the Status Tracking System.

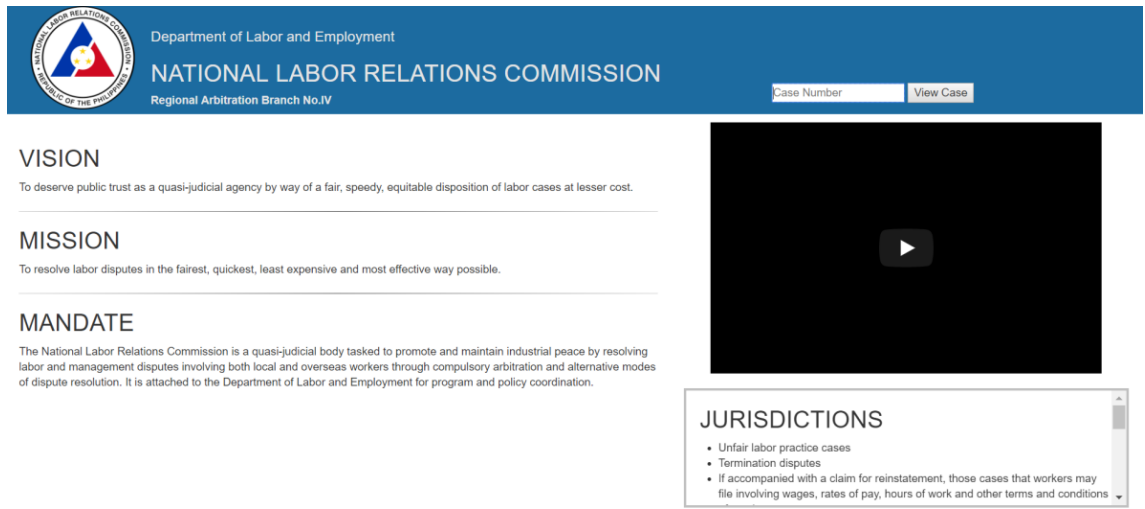

*Figure 5.* Case Status Tracking System for complainants and respondents

The status case tracking system provided a way of addressing issues of spending time, effort, and money by visiting the office for an inquiry on the status of the filed complaint. This tracking system provided information for people living far away from the Commission to be updated on their filed complaints with less effort.

***Data Privacy Act of 2012 of the Philippines Integrity Ensured on the Developed Web-based Management Information System***

The system used PHP's PASSWORD_HASH method to secure the passwords of all user accounts on the system. The PASSWORD_HASH method creates a 64-character hashed string that changes every time the same word is hashed. The PASSWORD_HASH method is the safest and most secure method of hashing as there are no ways to decrypt the method compared to the other hashing methods such as md5 and sha1 (PHP Manual, 2022).

The Exclusive OR (XOR) Cipher Algorithm was also used to apply information encryption to secure the records, specifically the names of both complainants and respondents. The XOR Cipher Algorithm is one of the cipher algorithms that is difficult to decrypt without knowing the key used in encryption, making this algorithm secure (101 Computing, 2020).

The method and algorithm were used to ensure data integrity and security in the system as per the Data Privacy Act of 2012 (Republic Act 10173, 2012) is shown in Algorithm 1.

| **ALGORITHM 1:** XOR Cipher Algorithm |
|---|
| *name* ← complainant's or respondent's name |
| *xorKey* ← encryption key |
| *nCtr* ← 0 |
| *charInName* ← every character in the *name* |
| *charInKey* ← every character in *key* |
| **for** *each* *nameChar* **in** *charInName*, **do** |
|     **for** *each* *keyChar* **in** *charInKey*, **do** |
|         *name[nCtr]* ← *nameChar* ^ *keyChar* |
|     **end** |
|     *nCtr* ← *nCtr* + 1 |
| **end** |
| return *name* |

***System Acceptability Using Standard Web Evaluation Criteria***

Its intended users and experts evaluated the developed web-based management information system of cases to measure the system's acceptability (Valaviius & Vipartien, 2013). The acceptability of the system was measured into three groups: quality, usability, and satisfaction models. The quality model was measured through accuracy, accessibility, appropriateness, efficiency, confidentiality, availability, portability, and recoverability. The usability model was measured through navigation and interactivity. The satisfaction model was measured through layout, information, and connection. The system's acceptability was evaluated as perceived by the intended end-users and experts in IT.

*IT experts.* The evaluation's overall result, as perceived by the experts, was 4.27, with a descriptive interpretation of "*Very Good*." It shows that the system provides quality data for the users and that the system will improve the daily operations of the Commission upon the implementation of the system. It also shows that the experts were satisfied with how the data was presented and displayed on the system but made suggestions and commented on the system's layout itself.

Table 2. Overall Summary of the Respondents' Ratings as Perceived by IT Experts to the Web-based Management Information System

| Item | Mean | Interpretation |
|---|---|---|
| Quality | 4.20 | Very Good |
| Usability | 4.42 | Very Good |
| Satisfaction | 4.19 | Very Good |
| **Overall Mean** | **4.27** | **Very Good** |

*End-users.* The system evaluation's overall result, as perceived by end-users, was 4.43, with a descriptive interpretation of "*Very Good*." It shows that end-users were satisfied with how data was presented in the system and how data was presented clearly, displaying what data was needed.

Table 3. Overall Summary of the Respondents' Ratings as Perceived by End-Users to the Web-based Management Information System

| Item | Mean | Interpretation |
|---|---|---|
| Quality | 4.42 | Very Good |
| Usability | 4.50 | Excellent |
| Satisfaction | 4.38 | Very Good |
| **Overall Mean** | **4.43** | **Very Good** |

## CONCLUSIONS AND RECOMMENDATIONS

In summary, the researcher has performed phases of the software development methodology to deliver the system's functionalities following the Commission's data. The system addressed the Commission's problems and issues in managing records of complaints, SEnA, and proper labor cases. The system ensured that the database's information was encrypted and was only accessible by a specific office. The system was evaluated by experts and its intended users and received a descriptive rating of *Very Good*, which means that the system performed its functionalities, presented information as needed, provided good navigation, and was accepted by both experts and its intended users.

The following recommendations were made, considering the findings and the conclusion of the study. They can be used by future researchers in the fields of information technology or information systems who want to develop the same kind of study. Recommendations for future development of the study are to (1) provide additional printable documents that include documents such as the mailing to both complainants and respondents; (2) complete the Agile software development methodology phases, continuing up to the release, tracking, and monitoring phase to finish the cycle; and (3) develop a further study that would identify the impact of using a web-based management system in a government agency.

## IMPLICATIONS

After the development of the web-based management information system for cases filed with NLRC RAB IV, the agency may improve its services to the public upon its implementation. With the status tracking system that may be utilized by any or both parties of respondents and complainants, a policy may be crafted to maximize the use of the status tracking system to reduce in-house inquiry of case status. Moreover, with the development of the different user levels within the agency, raffling of cases within labor arbiters will be more at ease, ensuring that the data being transferred is complete, correct, and precise.

## ACKNOWLEDGEMENT

This work received no funding upon its development. The researcher would like to thank Bulacan State University for the presentation of this paper at a national conference. This work cannot be done without the help of a friend, Mr. Jayson DC. Manuel, and the support and guidance of the author's colleagues, Dr. Digna S. Evale, Mr. Gabriel M. Galang, and Ms. Charlyn N. Villavicencio, from Bulacan State University.

**Author's Biography**

Aaron Paul M. Dela Rosa, MSIT, is currently a student in the Doctor of Information Technology program and currently working on his dissertation. Mr. Dela Rosa is a college instructor at the College of Information and Communications Technology (CICT) of Bulacan State University (BulSU). He is also an Online Learning Environment Specialist under the Educational Development Office (EDO) of the Office of the Vice President for Academic Affairs (OVPAA). He presents research papers at national and international conferences focusing on web application development and related fields. He is a member of multiple national and international organizations, and he is a certified data protection officer and certified in multiple programming languages.